\documentclass[twocolumn,showpacs,rsi]{revtex4}

\usepackage{graphicx}
\usepackage{dcolumn}
\usepackage{bm}
\usepackage[latin1]{inputenc}

\begin{document}
\title{The experimental realization of a two-dimensional colloidal model system}
\author{F. Ebert, P. Dillmann, G. Maret, P. Keim}
\affiliation{Fachbereich Physik, University of Konstanz, Box M621,
78457 Konstanz, Germany}
\date{\today}
\begin{abstract}
We present the technical details of an experimental method to
realize a model system for 2D phase transitions and the glass
transition. The system consists of several hundred thousand
colloidal super-paramagnetic particles confined by gravity at a flat
water-air interface of a pending water droplet where they are
subjected to Brownian motion. The dipolar pair potential and
therefore the system temperature is not only known precisely but
also directly and instantaneously controllable via an external
magnetic field $B$. In case of a one component system of
monodisperse particles the system can crystallize upon application
of $B$ whereas in a two component system it undergoes a glass
transition. Up to 10000 particles are observed by video microscopy
and image
processing provides their trajectories on all relative length and time scales.\\
The position of the interface is actively regulated thereby reducing
surface fluctuations to less than one micron and the setup
inclination is controlled to an accuracy of $\pm1\,\mu rad$. The
sample quality being necessary to enable the experimental
investigation of the 2D melting scenario, 2D crystallization, and
the 2D glass transition, is discussed.\\
\end{abstract}

\maketitle
\section{Introduction}
It is well known that dimensionality has a strong influence on the
macroscopic behavior of many physical systems. For example, the
\textit{Ising} model for ferromagnetics shows a phase transition for
2D and 3D but not \mbox{for 1D \cite{ising}.} Another example
concerning ordered phases where dimensionality plays a crucial role
is the existence of long-range translational invariance which exists
in 3D but not in 1D and 2D for finite temperatures. That energy
needed for a long wavelength deformation diverges in 3D for large
volumes but not in 1D and 2D. This enables thermal excitations to
destroy translational symmetry by long wavelength fluctuations
\cite{mermin1,mermin2}. A dynamical dependency on dimensionality $d$
is that the velocity autocorrelation function is dependent on delay
time $\tau$ like $\textbf{v}(\tau)\propto \tau^{-d/2}$. As the
diffusion constant is defined via the \textit{Green-Kubo} relation
$D=1/d \int_0^\infty d\tau\langle \textbf{v}(\tau)\textbf{v}(0)
\rangle$, the diffusion constant is finite in 3D but diverges in 2D
\cite{hansen}.\\
Colloidal model systems have proven extremely helpful to gain
insight into the fundamental mechanisms which govern solid state
physics. Here, we present a detailed description of the experimental
technique, sample preparation, sample properties of a specific 2D
colloidal model system ideally suited to study 2D physics.\\
The system consists of several hundred thousand colloidal
super-paramagnetic particles confined by gravity at a flat water-air
interface of a pending water droplet where they are subjected to
Brownian motion. The dipolar pair potential and therefore the system
temperature is not only known precisely but also directly and
instantaneously controllable via an external magnetic field $B$. In
case of a one component system of monodisperse particles the system
can crystallize upon application of $B$ whereas in a two component
system it undergoes a glass transition. Up to 10000 particles are
observed by video microscopy and image processing provides their
trajectories on all relative length and time scales.\\
Several questions of 2D solid state physics have been addressed
already using the system at hand. Some examples illustrating the
potential of the system are listed in the following.\\
The macroscopic melting behavior of crystalline systems sensitively
depends on the dimensionality. An intermediate phase exists in 2D
between fluid and crystal, the \textit{hexatic phase}: In this phase
the system has no translational order while the orientational
correlation is still quasi-long-range. Such a two step melting is
not known in 3D for isotropic pair interactions. The theoretical
melting scenario according to \textit{KTHNY}
\cite{hexatic_theo1,hexatic_theo2,hexatic_theo3} was successfully
confirmed experimentally using the system at hand
\cite{hexatic_exp,hexatic_exp2,7_1,7_2}. In particular, the
softening of the \textit{Youngs} modulus and \textit{Frank's}
constant predicted by renormalization group theory when approaching
the phase transition from the crystalline state was confirmed
experimentally \cite{9_1,9_2,9_3}. Furthermore, the direct control
of the system temperature by the magnetic field $B$ enables
ultra-fast quench measurements to investigate the growth and
time-development of 2D
crystals and glasses in out of equilibrium situations \cite{quench_patrick,quench_paper_glas}.\\
The possibility of introducing an anisotropic interaction potential
between particles by tilting $B$ off the normal of the 2D plane
allows for the investigation of the melting scenario of anisotropic crystals \cite{aniso_1,aniso_2}.\\
Introducing a second species of particles, i.e. using a binary
sample, it turned out that the system provides an ideal model system
for a glass former in 2D \cite{hansroland_glass}. In contrast to the
melting of crystals, the glass transition does not depend
characteristically on dimensionality as it exhibits the full range
of glass phenomenology known in 3D glass formers, both in dynamics
and structure \cite{bayer_2d_mct,epje_ebert}.\\
The system of binary dipoles shows \textit{partial clustering}
\cite{cluster_prl}, i.e. in equilibrium the smaller species
aggregates into loose clusters whereas the big particles are spread
more or less homogeneously. This heterogeneous local composition
leads to a variety of local crystalline structures upon supercooling
\cite{epje_ebert} causing frustration in the glassy state.
\section{Pending water drop geometry}
\begin{figure}[t]
\begin{center}
\resizebox{0.7\columnwidth}{!}{\includegraphics{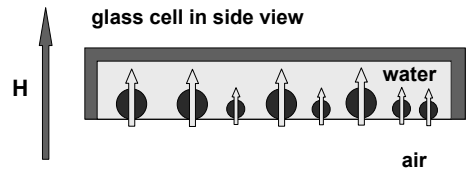}}
\end{center}
\caption[sq] {\label{cell}
  \em Super-paramagnetic colloidal particles (not to scale) confined at a water-air
  interface due to gravity. The curvature of the interface is
  actively controlled so that the surface becomes completely flat, and the system
  is considered to be ideally two dimensional. A magnetic field $\textbf{H}$ perpendicular to the interface
  induces a magnetic moment $\textbf{M}_i$ in each bead leading to a repulsive dipolar
  pair interaction.}
\end{figure}

The system described here consists of a suspension of one or two
kinds of micron-sized spherical super-paramagnetic colloidal
particles A and $B$ with different diameters $d_A$ and $d_B$ and
magnetic susceptibilities $\chi_A$ and $\chi_B$. Due to their high
mass density, they are confined by gravity to a water-air interface
formed by a pending water drop suspended by surface tension in a top
sealed cylindrical hole ($6\:mm$ diameter, $1\:mm$ depth) in a glass
plate. This basic setup is sketched in \mbox{Figure \ref{cell}}. A
magnetic field $\textbf{H}$ is applied perpendicularly to the
water-air interface inducing a magnetic moment $\textbf{M}= \chi
\textbf{H}$ in each
particle leading to a repulsive dipole-dipole pair interaction.\\
The set of particles is visualized by video microscopy from below
the sample and is recorded by an 8-bit CCD camera. The gray scale
image of the particles is then analyzed \textit{in situ} with a
computer. The field of view has a size of $\approx1\,mm^2$
containing typically $3\times10^{3}$ particles, whereas the whole
sample contains about up to $10^{5}$ particles. Standard image
processing provides size, number, and positions of the colloids.
Trajectories of all particles in the field of view can be recorded
over several days providing the whole phase space information. A
computer controlled syringe, driven by a micro stage, controls the
volume of the droplet to reach a completely flat and horizontal
surface. Thus, the ensemble is considered as ideally two
dimensional. Deviations from two-dimensionality are found to be
negligible: thermal excitations in vertical direction according to
the barometric height distribution are below $20\,nm$ for big
particles and below $100\,nm$ for the small particles. Capillary
waves as well as depths of dimples due to local
deformation of the interface are estimated to be below $20\,nm$.\\
Information on all relevant time and length scales is available, an
advantage compared to many other experimental systems. Furthermore,
the pair interaction is not only known, but can also be directly
controlled over a wide range. For all typical experimental particle
distances the dipolar interaction is absolutely dominant compared to
other interactions between particles like \textit{van der Waals}
forces or surface charges \cite{prl_gamma_zahn}.\\
The magnetic dipole-dipole pair interaction energy $E_{magn}$ is
compared to thermal energy $k_B T$ which generates \textit{Brownian}
motion. Thus, a dimensionless interaction parameter $\Gamma$ is
introduced by the ratio of potential versus thermal energy:
\begin{eqnarray}\label{glgamma}
\Gamma  & = &\frac{\mu_0}{4\pi}\cdot \frac{H^{2}\cdot(\pi
n)^{3/2}}{k_B
T}(\xi\cdot\chi_B+(1-\xi)\chi_A)^2\\
& \propto & \frac{1}{T_{sys}}.
\end{eqnarray}
Here, $\xi=N_B/(N_A+N_B)$ is the relative concentration of the small
species with $N_A$ big and $N_B$ small particles, $n$ is the area
density of all particles and $\mu_0$ is the permeability of vacuum.
For $\xi=1$ the interaction parameter reduces to that of a
one-component system \footnote{The definition of $\Gamma$ is
adjusted for reasons of tradition by several factors: A factor of
$\frac{1}{2}$ was omitted and $\pi^{3/2}$ is added. Setting
$r=n^{-1/2}$ implies a square arrangement of dipoles in the plane.
Other crystalline patterns or an amorphous arrangement would lead to
another prefactor. Here, it is not considered that the underlying
structure might change when altering the magnetic field, e.g. when
the system is undergoing a phase transition. Only the interaction
between two neighboring particles is taken into account although the
dipolar potential is long-range. Consideration of the interaction of
all particles leads to a \textit{Madelung} constant for a
crystalline pattern and a corresponding factor for an amorphous
arrangement. Another definition $\Gamma^\star= \frac{n^{3/2}(\chi_A
H)^{2}}{k_B T}$ for the interaction strength is often used where
only the big particles are considered and the significant smaller
contribution of the small
particles is neglected \cite{cluster_prl}.}.\\
Although the idea of the pending water droplet is simple, the
experimental realization is a technical challenge
\cite{zahn_diss,axel_diss,hans_diss,christoph_diss,peter_diss,flo_diss,patrick_diss}.
All difficulties that are discussed in this paper mainly result from
the subtle control of a flat water-air interface. Therefore the
question arises: What is the advantage of a free water-air interface
compared to a flat substrate which is far more easy to control? The
answer is: Because it provides absolutely uniform and free diffusion
in two dimensions for \textit{all} particles. This cannot be
guaranteed on a substrate. Uncontrolled interactions of particles
with the substrate, in particular pinning of at least a few
particles, is difficult to avoid. It turns out that many effects are
not visible when the system is placed on a substrate, e.g. the
continuous character of the phase transition from the crystalline to
the hexatic phase \cite{rice}.\\
A comparison of the system at hand with other 2D systems is given in
\cite{7_2}. The hardware and the software of the '2D colloidal
system' has been developed now for more than 15 years. The sample
quality of several different setups was significantly improved
during that time enabling progressively the access to new physical
questions. In the following an overview of the hardware and the
colloidal suspension is given, and the details to ensure high sample
quality are explained.
\section{\label{superpara}Colloidal suspension of super-paramagnetic
spheres}
\subsection{Super-paramagnetic particles} The colloidal
particles used for the system are commercially available
\cite{dynal}. They are porous polystyrene spheres doped with domains
of magnetite ($Fe_2O_3$) \cite{ugel}. The surface of the beads is
sealed with a thin layer of epoxy. The magnetic domains have a size
of typically ten nanometers, small enough for thermal energy to
overcome the magnetic coupling forces. Thus, magnetic moments are
distributed randomly. If no external magnetic field is applied, the
total magnetization is zero. Thus, the material exhibits no
remanence, a characteristic property of paramagnetic materials. The
prefix 'super' originates from the large susceptibility that is
comparable to that of ferromagnetic materials. Two types of
particles were used \footnote{Information on mass densities were
taken from the manufacturer as well as the diameter of the small
particles. The diameter of the big spheres was determined
microscopically by measuring the length of several hundred particle
chains in an in-plane magnetic field. Magnetic susceptibilities were
obtained by SQUID measurements (Group \textit{Prof. Schatz},
University of Konstanz) and they may vary between batches.}:
\begin{center}
\begin{tabular}{|c|c|c|}
  \hline
    species & \textbf{A }(big) & \textbf{B} (small)\\
  \hline
  diameter & $4.5\pm0.05 \:\mu m$ & $2.8 \:\mu m$\\
  \hline
   mass density & $1.5\: g/cm^3$ & $1.3\: g/cm^3$\\
  \hline
  susceptibility & $6.2\cdot10^{-11}\:Am^2/T$ & $6.6\cdot10^{-12}\:Am^2/T$\\
  \hline
\end{tabular}
\end{center}
Slices of particles observed with transmission electron microscopy
reveal that the big particles are quite monodisperse in size and
magnetic moment whereas the small particles might have higher
polydispersity \footnote{Big particles have $3\%$ polydispersity in
size (manufacturer information). No information is provided for the
small particles.}.
\subsection{Preparation of the colloidal suspension} The big particles are supplied by the manufacturer in
pure water solution, while the small particles are provided as
powder. To obtain a binary mixture with the desired relative
concentration $\xi$ of small particles and also the right absolute
concentration of particles, both suspensions have to be prepared and
characterized separately:\\
\textbf{Suspension of big particles:} The provided solution of big
particles is diluted with deionized water. To prevent aggregation of
the spheres, sodium dodecyle sulfate (SDS) is added until a
concentration of $c\approx0.9\times CMC$ is reached where $CMC$ is
the critical micelle concentration (CMC). SDS is an anionic
surfactant that covers the bead surface, with its polar end
directing towards the solvent away from the sphere. This sterically
stabilizes the colloidal suspension. Without SDS lots of particles
form aggregates due to \textit{van der Waals} attraction. To avoid
the growth of bacteria, the poisson Thimerosal ($1\,\mu l/ml$ of a
solution with $2\%$ content) is added. Sedimentation and aggregation
of the beads is avoided by storing the prepared suspension under
permanent rotation and weak ultrasonic treatment. It takes
approximately two days before the SDS has sufficiently stabilized
the colloidal particles and almost no aggregates are found in the
sample.\\
\textbf{Suspension of small particles:} The suspension of the small
particles is prepared by dissolving the powder in pure deionized
water and subsequent ultrasonic treatment. Thimerosal is added with
the same concentration as in the suspension of the big particles.
The
suspension of small particles is stable without adding any agents.\\
Both suspensions are directly mixed and the particle density was
measured microscopically to obtain the desired relative
concentration $\xi$.
\section{Experimental setup \label{setup_overview}}
\subsection{General hardware design}
The setup was constructed to measure extended 2D samples with the
option of manipulating them with light forces, i.e. fast scanned
optical tweezers \cite{Ashkin}. The details of the optical tweezers
are not described here as the focus of this work is only put on the
colloidal system. Nevertheless, it is necessary to mention their
implementation to understand the general design of the experimental
setup. Optical tweezers basically work as follows: a focused laser
beam exerts light forces to an object that has a different index of
refraction with respect to the surrounding fluid. In the experiment
at hand, colloidal particles can be trapped in the
focus of a laser beam.\\
The separation of tweezers and microscope objective allows an
optimum choice of objectives for each task: optical tweezers ideally
have an objective with high numerical aperture and high
magnification, while a microscope objective with small magnification
is suitable to observe a large field of view. The optical tweezers
have to access the sample from top, and therefore the microscope has
to be placed below the sample. This configuration is necessary
because light pressure exerts a considerable force onto the
particles in beam direction. This force is compensated by surface
tension while under tweezers illumination from below the particles
are pushed upwards.\\
\begin{figure}[t]
\begin{center}
\vspace{1cm}
   \includegraphics[width=1.0\columnwidth]{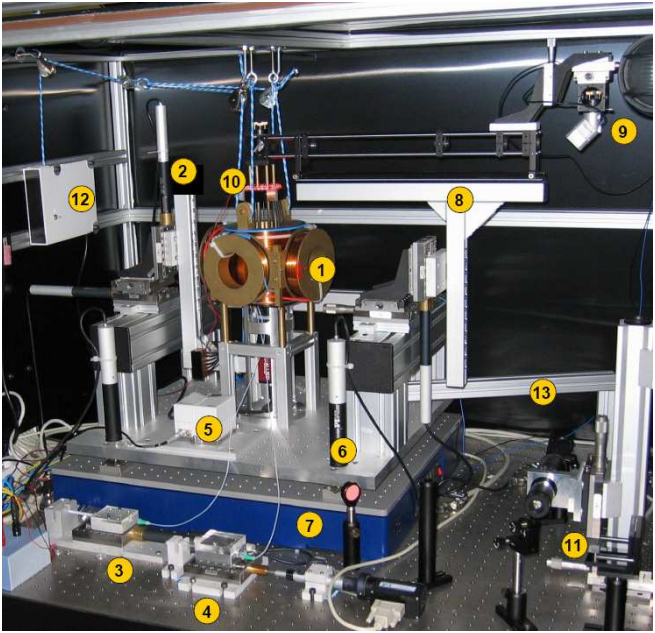}
\end{center}
\caption[sq] {\label{setup} \em Experimental setup with optical
tweezers to prepare a 2D monolayer of super-paramagnetic particles
at a water-air interface. The indicated parts are enumerated and
explained in the main text.}
\end{figure}
The experimental setup is shown in Figure \ref{setup}. The main
components are described, following their labeling in Figure
\ref{setup}:
\begin{enumerate}
\item Five \textbf{copper coils} generate the external magnetic
field with controllable xyz-components. The monolayer lies in the
xy-plane, and the z-direction is perpendicular to it. The sample
position is in the center of the large coil that generates the
interaction potential between the particles. Lateral coils
compensate the in-plane component of the Earths magnetic field.
These coils can also be used to tilt the total magnetic field with
respect to the samples normal. In this way an anisotropic dipole
interaction in the sample plane can be achieved
\cite{christoph_diss}. The coils are wounded layer by layer and
provide a very homogeneous field in the volume where the sample is
located. In xy-direction the field increases from the center towards
the inner side of the central coil. In z-direction the field
decreases away from the center. The deviations from constant field
in the area of interest ($6\times6\,mm^2$)
are smaller than $0.5\%$ as measured with a hall sensor.\\
The magnetic field is kept constant to fluctuations of less than
$0.5\,\mu T$ by a user-specific designed constant current source. It
compensates the change in electric resistance which results from the
heating of the coils by the electric current itself. The actual
current is measured with a current digital meter
\footnote{\textit{Keithly Instruments}, 2700
Multimeter Integra series.} to calculate $\Gamma$ from gauge with the hall sensor.\\
The sample holder and sample cell are obscured by the coils and are
explained separately in section \ref{holder}.
\item The \textbf{microscope optics mount} is held by three linear
positioning stages \footnote{\textit{Newport}, M-UMR8.25 with
$25\,mm$ travel range. All actuator driven positioning stages are of
this type.} that are driven by computer controlled linear actuators
\footnote{\textit{Physik Instrumente}, DC Mike 230.25 with minimum
incremental steps of $50\,nm$ and travel range $25\,mm$. All
actuators in the setup are of this type if not explicitly specified
otherwise.}. The microscope optics consists of a $4\times$
microscope objective, an optical tube with magnification $1\times$
and a gray scale 8-bit CCD camera. Further, the \textbf{light source
No. I} consisting of 24 LEDs is mounted with light guides leading
directly underneath the sample (for details see section
\ref{holder}).
\item The \textbf{sample water supply actuator} drives a
  conventional $1\,ml$ syringe filled with deionized water. A
teflon hose connects the syringe directly with the colloidal
suspension in the glass cell. The exact amount of water and
therefore the curvature of the water-air interface is adjusted
directly and computer controlled by this actuator.
  \item The \textbf{water basin actuator} controls the amount of water in the water
pocket underneath the sample. It is used to keep the atmosphere in
the sample chamber at constant humidity.
  \item The \textbf{\textit{Nivel} inclination sensor} \footnote{\textit{Leica Geosystems AG}, Nivel20.} is mounted on the experimental plate
and measures the inclination of the whole setup with an accuracy of
$\pm\,1\,\mu rad$. As the sample is very sensitive to changes in
tilt with respect to the horizontal, this accuracy is necessary to
ensure a sufficient absolute positioning of the whole setup.
Temperature variations or manipulation by the experimentalist are
causing severe deviations in tilt which need to be compensated. The
tilt control is specified in section \ref{tilt_control}.
  \item The experimental plate is adjusted by two \textbf{heavy duty
actuators}\footnote{\textit{Physik Instrumente} DC Mike 235.5DG.}
forming two stands of a tripod. The third stand is a static spike
below the plate located a few centimeters behind the coils away from
the actuators. Together with the inclination signal of the
\textit{Nivel} sensor the computer controlled actuators ensure
precise positioning of the whole experimental plate. Slow deviations
as from thermal expansion are compensated (see section
\ref{tilt_control}).
  \item A \textbf{piezo table} \footnote{\textit{HWL Scientific}, TS150.} is a
dynamic vibration isolation system suppressing fast vibrations
$>\,1\,Hz$ like foot fall sound or building vibrations. The large
and massive table, where the experiment is located, stands on rigid
pillars. Air damping is switched off as the piezo table is only
working sufficiently well when placed on a rigid underground.
  \item The \textbf{optical tweezers mount} carries a micro bench with two
lenses that conjugate the plane of the piezo deflector to the
tweezers objective (obscured by coils). At the right end of the
micro-bench the piezo deflector is mounted. The tweezers objective
mount is attached at the left end.
\item The \textbf{piezo deflector} \footnote{\textit{Physik Instrumente}, Scanner: S-334 2SL, Wavegenerator: E516.}
is mounted in $90°$ geometry to deflect the vertical IR laser beam
\footnote{\textit{Spectra Physics}, Millennia IR, diode pumped solid
state YAG laser, output wavelength $\lambda=1064\,nm$, max. power
$10\,W$.} towards the optical axis of the micro bench. The exact
position of the scanner is manually adjusted by three linear stages
and a tilt platform in two directions. The computer controlled
scanner enables fast manipulation of the focus position inside the
sample plane. The scanner can be replaced by a much faster
acousto-optical deflector (AOD).
  \item The \textbf{tweezers objective mount} is connected to the
optical micro bench and deflects the laser beam towards the sample.
Further, the \mbox{\textbf{light source No. II}} is attached (copper
mount) carrying 24 LEDs. The light is guided beside the tweezers
objective into the sample via light guides. To prevent thermal
heating, a minimum distance of $10-20\,cm$ between sample and LEDs
has to be assured.
  \item The IR laser beam is guided by the \textbf{laser optics} to the deflector.
Here, IR light is used due to lower absorption of the particles
compared to visible light. Absorption weakens the laser trap
stiffness as light pressure increases compared to gradient forces of
the electric field.
Furthermore, heating of the surrounding solvent causes local convection.\\
A beam expander broadens the beam diameter to exploit the full
diameter of the mirror or the aperture of the AOD respectively which
increases the trap quality.
  \item The \textbf{crane} with pulley is used to lift the coils. To
exchange the sample inside the sample holder, the coils
($\approx20\,kg$) have to be lifted to access the glass cell.
  \item The \textbf{camera fan} reduces the heat emitted by the camera to minimize
thermal disturbance of the sample.
\end{enumerate}
\subsection{Sample holder and microscope optics\label{holder}}
\begin{figure}[t!]
\begin{center}
\includegraphics[width=0.95\linewidth]{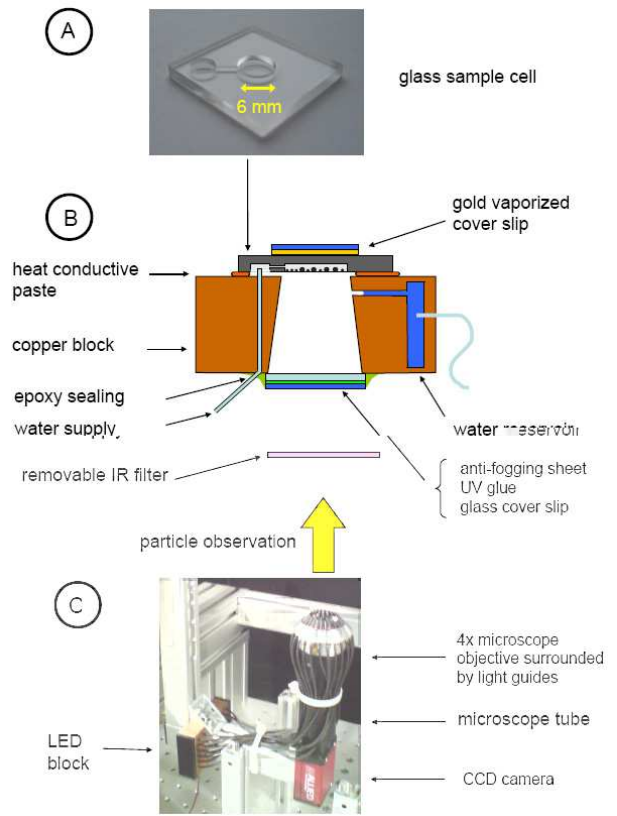}
\end{center}
\caption[sq] {\label{block} \em A: Picture of the glass sample cell,
upside down. The rim of the center bore is where the water-air
interface is attached. The bore is connected by a channel to a side
bore that serves as a reservoir for the water supply nozzle. B:
Schematic drawing of the glass cell holder with composite bottom
window to illuminate and observe the sample. C: The microscope
optics mount with wave guides is directly located below cell holder
(A). The camera with microscope tube and objective is surrounded by
fibre optics that guide light from 24 LEDs ($\lambda=624\,nm$,
located in heat sink copper block) to the sample (light source No.
I). The sample is illuminated with diffusive light. Here, sample
holder and magnetic field coils are dismounted.}
\end{figure}
\begin{figure}[t!]
\begin{center}
\includegraphics[width=0.95\linewidth]{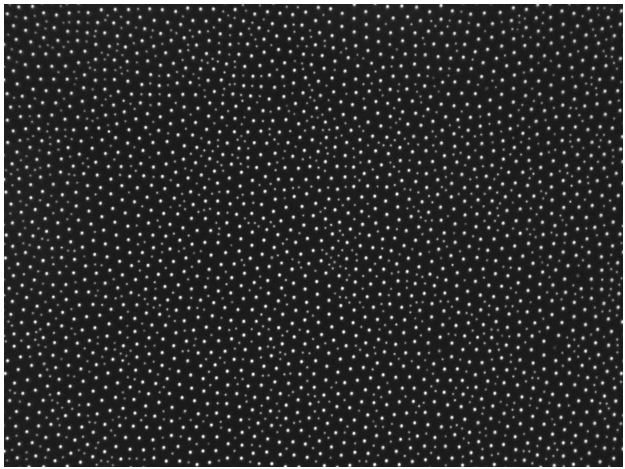}
\end{center}
\caption[sq] {\label{raw} \em Raw image of 1613 big and 1159 small
particles ($\xi=42\%$) in the field of view of $1158 \times 865 \:
\mu m^2$ in diffusive illumination (light source No. I, 8-bit gray
scale CCD camera, $4\times$ microscope objective). The particles
were confined at a flat and horizontal water-air interface. An
external magnetic field of $B=4.70\,mT$ is applied corresponding to
an interaction strength of $\Gamma=660$. Big and small particles can
be clearly distinguished by their apparent size.}
\end{figure}

In Figure \ref{block} the sample cell, sample holder and light
source are shown in detail. The construction of the interior is very
subtle as small changes might influence the sample behavior
drastically via illumination, thermal gradients and atmosphere
inside the sample chamber. The sample is observed from below to
enable access of the optical tweezers from top. For a clear
observation the bottom of the chamber has to be transparent. At the
same time, the chambers atmosphere has to be saturated with water
vapor to minimize the evaporation from the sample cell. That,
however, causes fogging on ordinary transparent windows based on
silicon dioxide ($SiO_2$) or conventional polymers like
Polymethylmetacrylat (PMMA). The solution of this problem is
presented in the following where the sample holder geometry is explained.\\
The glass sample cell \footnote{\textit{Helma}, purpose-built glass
cell.} is shown in Figure \ref{block}A. The center bore contains the
colloidal suspension which is held by surface tension at the sharp
edges of the cylindrical bore. To enhance wetting contrast, the flat
area outside the bores are treated with silane
\footnote{\textit{Amersham}, PlusOne Repel-Silane ES.} making this
area water repellant. The bores are treated afterwards with $20\%$
RBS solution \footnote{\textit{Roth}, RBS35 Konzentrat.} to ensure
they are hydrophilic. The small bore is accessed by the nozzle of a
teflon hose (see Figure \ref{block}B) to control the amount of water
in both bores. The curvature of the
monolayer is thereby controlled directly with the water supply actuator.\\
Figure \ref{block}B gives schematic insight into the sample holders
geometry. The holder is made from massive copper to provide a
sufficient heat sink being unsusceptible to quick temperature
variations. For thermal contact of the glass sample cell heat
conductive paste is used. Additionally, this seals the chamber
against evaporation of water. From below the chamber is closed by a
composite window that consists of a conventional cover slip glued
with an anti-fogging sheet \footnote{\textit{PINLOCK},
www.pinlock.nl (\today), a motorcycle helmet shield of
$\approx1\,mm$ thickness.} by transparent UV glue
\footnote{\textit{Norland Products}, Norland optical adhesive 61
Lot226.}. It is sealed to the copper block with epoxy glue. This
window provides three necessary properties: (i) anti-fogging
behavior, (ii) clear transparency for observation, and (iii)
impermeability for water vapor. It was found that the impermeability
for water vapor is a crucial point. A strong evaporation was always
correlated with strong particle drift (up to $0.5\,mm/day$), at
least in a binary mixture. A data acquisition without particle drift
was only possible using the composite window. Furthermore, the
lifetime of the sample is limited by water volume of the syringe
which is depleted after
approximately half a year for a high evaporation rate.\\
Additionally to the composite window a water basin in a side pocket
of the sample holder is necessary to saturate the atmosphere.\\
The gold platelet \footnote{100\,nm gold layer vaporized on a
conventional coverslip.} reflects light back into the sample cell.
With the light source No. I (below sample) this platelet is
necessary for sufficient illumination. Further, the light is
reflected by the inner walls of the copper block which makes this
particle illumination sensitive to water condensation.\\
Figure \ref{block}C shows the microscope optics. A gray scale 8-bit
CCD camera \footnote{\textit{Allied Vision Technology}, Firewire
camera Marlin 145B.} is connected via a microscope tube
\footnote{\textit{Stemmer Imaging}, c-mount microscope tube with
magnification $1\times$.} to a microscope objective
\footnote{\textit{Olympus}, UIS2 series PLN4X, 0.10 numerical
aperture, $18.5\,mm$ working distance.}. The sample holder and coils
are dismounted. The microscope objective has a working distance of
$18.5\,mm$ and is located directly underneath the sample chamber. In
this arrangement there is enough space to turn an IR filter
\footnote{\textit{Edmund Optics}, TechSpec Shortpass Filter - 850NM
25mm Dia.} between objective and composite window to block the laser
tweezers beam (which is focused on the CCD chip of the camera as the
laser focus is in the observed particle plane). The light source No.
I consists of 24 LEDs \footnote{LEDs $\lambda=624\,nm$ with narrow
beam divergence $\theta=6°$ and brightness $10000\,mCd$.} placed in
a copper block as heat sink. To avoid thermal disturbance, the
diodes are located far away from the colloidal suspension. Light
guides \footnote{\textit{Goodfellow}, PMMA fiber wave guides, fiber
diameter $1\,mm$.} lead towards the sample with the ends located
around the objective and point directly at the colloidal monolayer
from below.\\
A microscope image of the particles is imported with a repetition
rate of $10\,Hz$ via \textit{firewire} connection to a computer for
further processing. A typical image obtained with this optics is
shown in Figure \ref{raw} with approximately 3000 particles
\footnote{In a one-component sample, densities with up to 10000
particles in the field of view can be prepared using this optics. In
the binary case, the distinction between particle species becomes
difficult for more than $\approx4000$ particles.}. The field of view
has a size of $1158 \times 865 \: \mu m^2$ with a resolution of
$1392 \times 1040 \,Pixel$. The diffusive illumination (light source
No. I) provides a clear contrast, and particle species can easily be
distinguished.
Both light sources No. I and No. II have different advantages and
disadvantages: Light source No. II provides a more homogeneous
illumination over the whole sample than light source No. I
illuminating from below. The advantage of LED source No. I is that
data can be recorded right at the edge of the sample cell. This is
of interest for investigations of the monolayer close to a hard
wall. At the edge, other illumination techniques fail, like
classical \textit{Köhler} illumination or illumination with light
source No. II. There, light is strongly scattered at the edge of the
cell.
\subsection{Image processing\label{image}}
Raw images have to be processed \textit{in situ} for two reasons:
Firstly, the stabilization mechanisms described in section
\ref{stabilisation} require present information, like mean particle
size, particle density, and coordinates of all particles to control
the system. Secondly, storing raw images for later image processing
would exceed the storage capacity by far. Even the processed data
exceeds the storage capacity if particle coordinates are recorded in
equal time intervals of e.g. $\Delta \tau=1\,sec$. Therefore,
particles are tracked \textit{in situ} and a 'multiple $\tau$'
algorithm can be used to increase the time steps between stored data
snapshots (see section \ref{multiple}). Thus, image processing and
\textit{in situ} tracking of coordinates have to be fast ensuring
rapid data acquisition and unhindered sample control. The steps of
image processing are now explained: \\
\textbf{Raw data image:} An 8-bit gray-scale image is imported from
the CCD
camera.\\
\textbf{Binary image:} The image is converted to a binary
black/white image by setting the pixel values to one if the
intensity is above a certain threshold value (cutoff) and zero
elsewhere. Beside the large areas resulting from big and small
particles ('blobs'), small
noise artifacts are still present.\\
\textbf{Erosion - dilation:} Noise artifacts have to be removed by
eroding and subsequent dilation of the blobs by a layer of one pixel
thickness. Small blobs like single pixels or pixel chains vanish.\\
Blob labeling: In the next step, connected pixels are assigned with
particle labels ranging from one to the number of found blobs.\\
\textbf{Blob size histogram:} In a binary sample a histogram of the
sizes of these labeled areas is produced. A clear discrimination of
two blob sizes is possible due to the sharpness of the average blob
sizes. In a binary sample, this discrimination restricts the
particle density of the sample to a maximum of $\approx4000$
particles in the field of view, else a clear assignment of the
particle species is difficult. A chosen separator value is used to
divide the histogram in two parts. Each of these parts is averaged
to obtain the mean blob size of both particle species, and the
integral over each part
provides the number of particles of each species.\\
\textbf{Coordinates from center of mass:} Finally, the coordinates
of the particles are determined by calculating the center of mass of
each blob. The labels of all data sets are then synchronized in time
to obtain trajectories.\\
Summary of the input/output parameters: Input are a 8-bit gray-scale
image with particle features, a cutoff value for intensity and a
separator value to discriminate particle species if the sample is
binary. The image processing provides as output: coordinates of each
blob, particle species, mean blob size for each species, and the
total number of particles for each species in the field of view,
$N_A$ and $N_B$. For the data acquisition, each snapshot provides a
$(N\times4)$ floating point data array with columns $(x,y,t,l)$
where $x$ and $y$ are coordinates, $t$ is the time-step, and $l$ are
the labels of each particle. The species of each particle is coded
in the label column with the sign of the label value to save storage
volume. Big particles are labeled positive and small particles
negative.\\
\subsection{Software control \label{software}} All integrated devices
like camera, actuators, inclination sensor, laser scanner, constant
current source, magnetic field hall sensor, or IR laser are
simultaneous controlled by a single computer software programmed in
\textit{Interactive Data Language (IDL)} \footnote{\textit{ITT
Visual Information Solutions}, http://www.ittvis.com (\today).}.
Running for several years
with almost no pause, the control software worked stable without exception.\\
\textit{IDL} provides a universal platform, not only for controlling
the experiment, but also for data acquisition and data evaluation.\\
When stabilizing the sample over many months, it is desirable to
have full access and information of the system 24 hours a day.
Especially when control parameters of the regulation mechanisms are
unstable, the manual input of the experimentalist is necessary. To
ensure permanent supervision, the whole experiment is controlled via
a single computer \footnote{\textit{Intel} Pentium IV $3.4\,GHz$,
$1\,GB$ RAM.}. If a control parameter is out of range, the computer
control contacts the experimentalist via email. Then, a text message
is generated by the email account to inform the experimentalist via
mobile phone. The experiment can then be fully controlled via a
remote control program from every computer with internet connection.
Also data acquisition can be started and stopped by remote control.
\section{Stabilization of the monolayer and sample quality\label{stabilisation}}
For stabilization and equilibration of the sample, it is necessary
to keep system parameters constant against perturbations.
Furthermore, controlled and fast changes of parameters have to be
applied, without over- or undershooting the desired value. In
general, feed back loops are used to perform such tasks.\\
In the 2D colloid experiment of this work, several interacting
feedback loops are used to ensure system stability: 1) Water supply
control of the water-air interface; 2) Particle density control in
the field of view; 3) Tilt control of the whole setup; 4)
xy-position of the camera for compensating the sample drift; 5)
Current control for the magnetic field coils; 6) Position of all
actuators, the piezo scanner
and the damping table.\\
Especially the regulation mechanisms 1), 2), and 3) are heavily
interacting. A typical scenario is the following: The tilt of the
whole setup is changed by the tilt regulation to adjust an
asymmetric density profile. The system will respond with a drift of
particles to equilibrate over typically one day. This changes the
number of particles in the field of view, which is then compensated
by the area density regulation that is directly coupled to the water
supply control.\\
Therefore, the parameters of the different controls have to be
adjusted carefully with respect to the characteristic timescales of
the regulations.\\
The detailed regulation mechanisms of the experiment and the
adaption to the standard \textit{Proportional-Integral-Differential}
(PID) control is described in the following \footnote{The necessary
nomenclature of a PID feed back loop is explained using a typical
application, a thermostat: the three \textit{PID} parts take into
account the \textit{actual value} (temperature at the moment) and
the history of the \textit{process variable} (temperature history)
and add up to a \textit{correction variable} (heating power) which
is applied to the system in order to reach the
\textit{set-point} (desired temperature).}.\\
\subsection{\label{water_supply}Water supply control}
\begin{figure}[t]
\begin{center}
\includegraphics[width=\columnwidth]{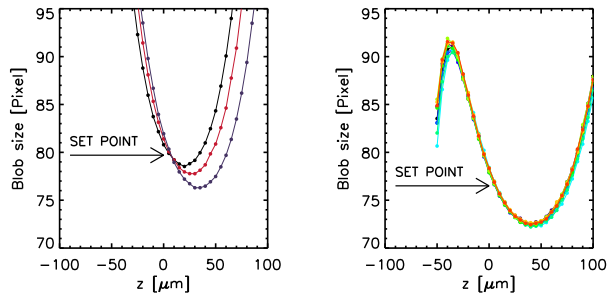}
\end{center}
\caption[sq] {\label{zscan} \em Blob size depending on the distance
between the focus plane and the particles. During regulation the
focus of the objective is always at $z=0\,\mu m$, and the z-position
of the particles is at the minimum of the z-scan. Left: Three
z-scans of a sample where the particle density was lowered over four
hours from $N_A=2300$ to $N_A=2100$ big particles in the field of
view. The apparent size of the particles is obviously strongly
dependent on the area density of the sample due to mutual
illumination of the particles (z-scans shift to lower values and to
the right over time). The shift to the right is caused by the water
supply control that fixes the \textit{set-point} of the blob size.
Therefore, all curves have to intersect at $z=0$. Right: Ten z-scans
(two hours waiting time between two scans) of a stable sample with
constant density. The focus plane is inside the water in a distance
of $42\,\mu m$ relative to the monolayer.}
\end{figure}
The water-air interface is kept at a fixed height by regulating the
water volume in the glass cell with a computer controlled nanoliter
pump. The position of the interface relative to the focus of the
observing microscope objective is obtained from the apparent blob
size of the big particles. This is demonstrated in Figure
\ref{zscan}. The position of the objective is scanned over a range
of $\pm 100\,\mu m$ in vertical direction, and the apparent blob
size is changing by $\approx 25\%$ (this value is also dependent on
other parameters from image processing described in section
\ref{image}). Setting the focus position inside the water above the
particles, enables the water supply control to detect a relative
change of the interface height by a change of the blob size.
Subsequently, this change can be
compensated by adjusting the water volume in the cell.\\
The particle density plays a crucial role in the regulation of the
interface height, because in diffusive light geometry the particles
mutually illuminate each other by reflection. Thus, the apparent
blob size is decreased when the particle density is lowered and vice
versa. The water supply control cannot distinguish this effect from
a real height change of the interface as it fixes the
\textit{set-point} of the blob size. The consequence is a shift of
the z-scan as shown in the left graph of Figure \ref{zscan}. There,
the particle density was lowered by $\approx 10\%$.  The shift of
the z-scan is an accompanying effect of the particle density
regulation (see chapter \ref{densityreg}). The left side (microscope
focus inside water) of the parabolic z-scan was chosen for
regulation because a perturbation of the particle density in any
direction is damped by this illumination effect \footnote{Assume a
perturbation of the particle density where the density is lowered:
1) as a consequence, particles illuminate each other less, 2) the
apparent blob size decreases, 3) the water supply control reacts, as
if the interface height was shifted upwards: water is pumped into
the cell. 4) this counteracts the original perturbation, because
particle density is increased by this. A perturbation towards higher
densities is analogously counteracted. Regulation at the other side
of the z-scan (focus below interface) is not advisable as it has the
opposite effect, a reinforcement of perturbations.}. It additionally
stabilizes the regulation compared to \textit{Köhler}
\mbox{illumination \cite{peter_diss,hans_diss}}, where this effect is not found.\\
\begin{figure}[t!]
\begin{center}
\includegraphics[width=\columnwidth]{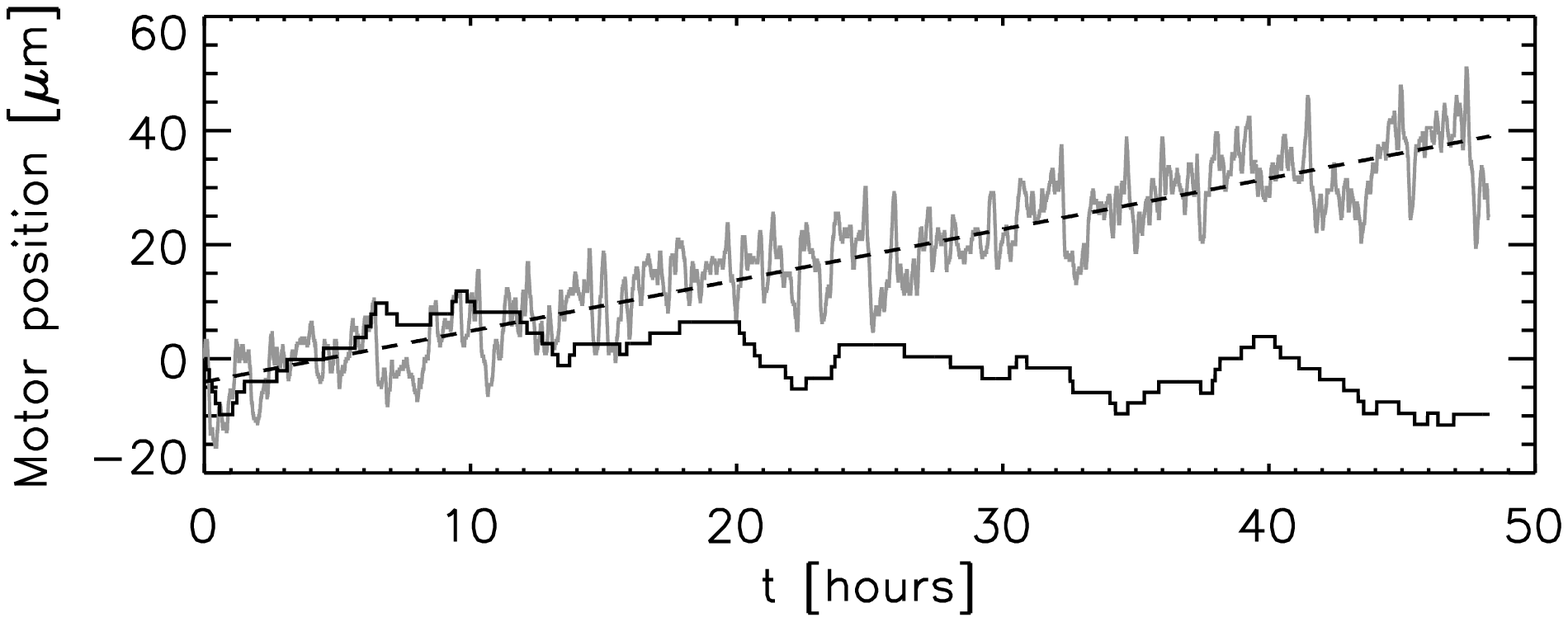}
\end{center}
\caption[sq] {\label{pos_spritze_cam} \em Position of the camera
motor (black, plotted with factor 10) and the syringe motor (grey).
The fit (dashed line) shows a continuous increase of $0.89\,\mu
m/hour$ of the syringe motor due to evaporation of water from the
water-air interface.}
\end{figure}
\begin{figure}[t!]
\begin{center}
\includegraphics[width=\columnwidth]{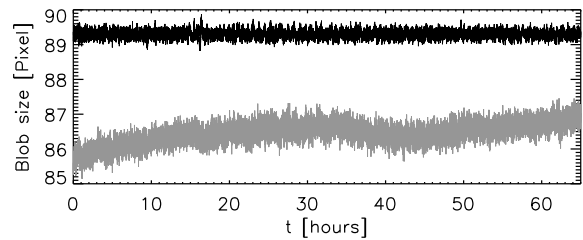}
\end{center}
\caption[sq] {\label{blobsize} \em Blob sizes of the small (grey,
shifted by $+40$ Pixels) and the big particles (black). The blob
size regulation keeps the apparent size of the big particles
constant beside fluctuations of $\pm 0.15$ Pixels (FWHM/2).
According to the slope in the z-scan this corresponds to a
fluctuation of $\,\approx 1\,\mu m$ in the distance between focus
and particle plane. The interaction strength was $\Gamma=423$.}
\end{figure}
The curves of the scans are stable in time when the density in the
field of view is constant. No significant change is seen in the
right graph of Figure \ref{zscan} over $20$ hours, where the sample was in equilibrium.\\
A regular \textit{PID} regulation is not advisable for the water
supply control because the deviation of the \textit{process
variable} (blob size) is not only dependent on the interface height
but also on the choice of the focus position, the illumination
properties, the particle density, the relative concentration of big
and small particles, and the parameters of the image processing
(cutoff, separator). All these parameters change the z-scan and
thereby the slope at the \textit{set-point} at position $z=0$. A
feed back loop that is much less sensitive to variations of these
parameters is a simplified three-step \textit{proportional} term.
The deviation around the \textit{set-point} is divided in three
parts: (i) A range around the \textit{set-point} where no water is
pumped and the \textit{set-point} is considered to be reached, (ii)
a range where a constant quantity of water is pumped, and (iii) a
range where this amount is quadrupled. The \textit{correction
variable} of the water supply control is the position of the water
pump actuator plotted in Figure \ref{pos_spritze_cam} as the grey
curve. A continuous increase is found resulting from water
evaporation of the interface. The width of the deviations can be
traced back to a backlash in the syringe where a rubber piston is
pushing the water. This three-step proportional feedback loop holds
the \textit{set-point} value of the average particle size at
$(89.3\,\pm\,0.15)\,Pixel$ in the example shown in Figure
\ref{blobsize}. The fluctuation is the FWHM of the deviation from
the \textit{set-point} value and is less than $\pm 0.2\,\%$ for 60
hours. Thus, the deviations of the interface position relative to
the observation objective are less than $1\,\mu m$ according to the
slope in the \textit{set-point} of the z-scan in Figure \ref{zscan}
(the slope at $z=0$ is used to obtain the fluctuations of the
interface height from the fluctuations of the blob size). However,
in this estimation it is assumed that the z-scan is absolutely
constant which is not generally true. The apparent size of the small
particles can fluctuate slowly (here, less than $3\%$ in 60 hours).
The reason in this particular case is suspected in a slight change
of illumination intensity over days due to condensation of water at
the walls inside the sample chamber. The particle density control,
as explained in the following, assures that this does not lead to a
perturbation of
the interfaces curvature.\\
\subsection{\label{densityreg}Particle density control}
\begin{figure}[t]
\begin{center}
\includegraphics[width=\columnwidth]{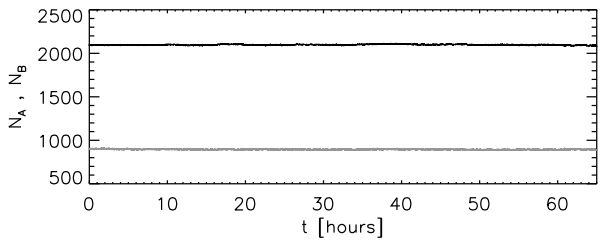}
\end{center}
\caption[sq] {\label{density} \em Number of the $N_A$ big (black)
and $N_B$ small particles (grey) in the field of view are plotted
for a time interval of almost three days at $\Gamma=423$. The
fluctuation of the small particles during the observed time is
$\pm1.3\%$ (FWHM/2) and for the big particles $\pm0.6\%$ (FWHM/2).}
\end{figure}
To control the particle density in the field of view, the curvature
of the interface has to be changed. By pumping water in the droplet
the curvature is increased and the particle density in the center of
the sample is raised due to downhill slope forces.\\ The focal plane
of the observation objective is fixed at a constant distance to the
particle interface by the water supply regulation as described in
the previous section. This means that a change in the microscope
objective position is followed by a change of the same distance in
the particle interface position. The focal plane position is
therefore used as the \textit{correction variable} for the
particle density in the field of view.\\
To avoid resonant feed back of the control loops, a prerequisite for
the density regulation are separate timescales of both regulations:
the timescale of the water supply regulation is much shorter
compared to that of the density regulation (minutes compared to
hours). The flow of particles reacts very slowly when the curvature
is slightly
changed, whereas the interface height is changing instantaneously upon a change in water volume.\\
The feed back loop is chosen as a \textit{proportional-differential}
control with negative \textit{reset time} (i.e.
\textit{differential} term is damping) and no \textit{integral} term
(overall particle number is a conservative \textit{process
variable}). To reduce the influence of noise to the
\textit{correction variable}, two damping mechanisms are introduced:
Firstly, the derivative of the particle density is averaged over an
elapsed time of \mbox{$\approx 0.5$ hours}. Secondly, the
\textit{correction variable} is added up until it reaches a
threshold before the z-actuator is driven. Only the number of big
particles $N_A$ is used as the \textit{process variable}. In case of
a binary mixture, the number of small particles is thus indirectly
regulated. Using the sum of both species as \textit{process
variable} leads to instabilities when the sample changes its
relative concentration in the field of view. The position of the
z-actuator, the \textit{correction variable} of the particle density
regulation, is plotted in Figure \ref{pos_spritze_cam} over 48
hours. Fluctuations are in the range of $\pm1\,\mu m$. This is an
upper estimate for the long-time fluctuation of the interface
position relative to the glass cells edges. The real height
fluctuations are expected to be less as the particle density control
additionally compensates deviations that originate from the water
supply control as explained above (e.g. illumination effects, see
section \ref{water_supply}). Thus, the correction via the focus
position (black curve in Figure \ref{pos_spritze_cam}) is only
partially necessary to compensate a deviation of the interface
height.\\ The accuracy of the density control is demonstrated in
Figure \ref{density} for both particle species in equilibrium at
$\Gamma=423$. The fluctuation of the small particles during the
observed time is $\pm1.3\%$ (FWHM/2, small particles) and for the
big particles $\pm0.6\%$ (FWHM/2, big particles). The fraction of
big particles at the edges of the field of view is $\approx 9\%$ and
of the small particle $\approx 13\%$. These particles are likely to
drop in and out of the field of view by their thermal motion and
contribute to the measured density fluctuations. Thus,
the real area density fluctuations in the field of view are expected to be even smaller.\\
The constant area density of both particle species reflects a main
aspect of the sample quality.
\subsection{Setup tilt control\label{tilt_control}}
\begin{figure}[t!]
\begin{center}
\includegraphics[width=\columnwidth]{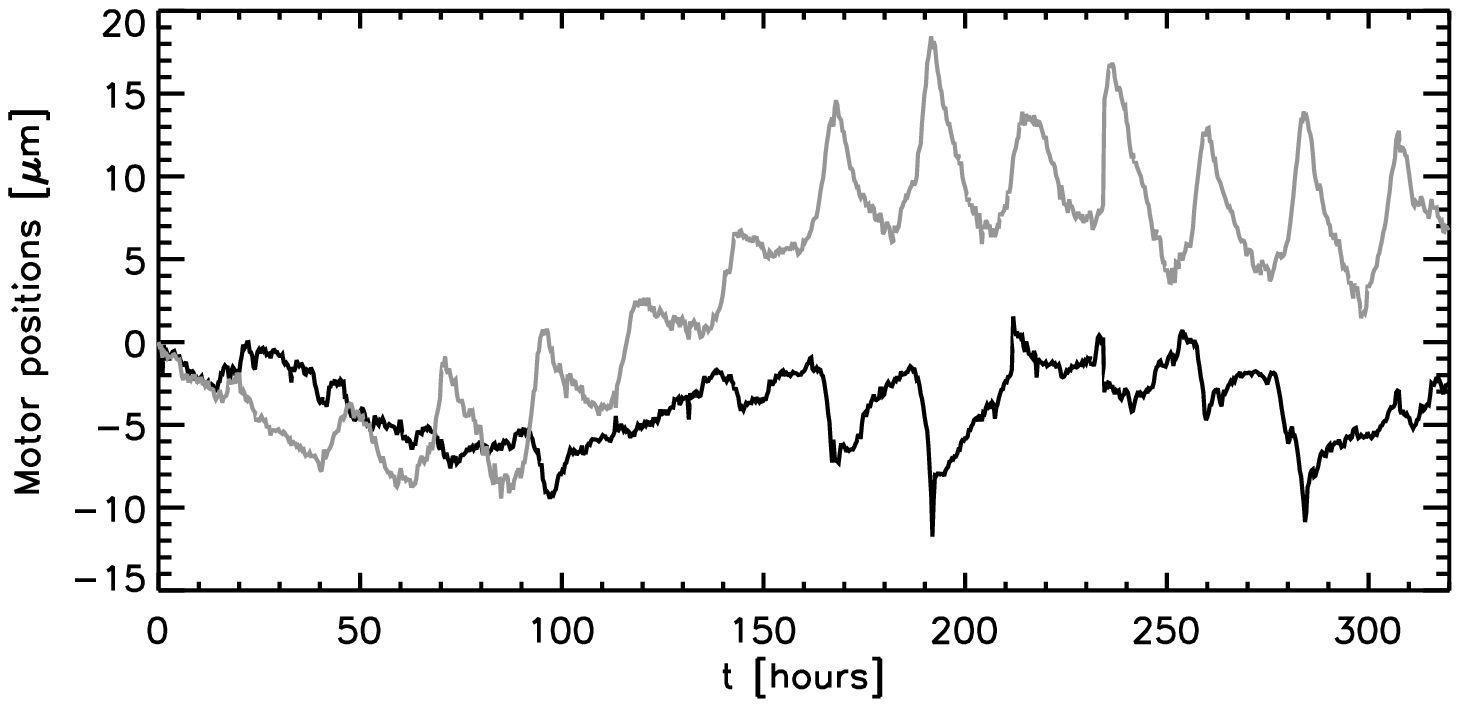}
\includegraphics[width=\columnwidth]{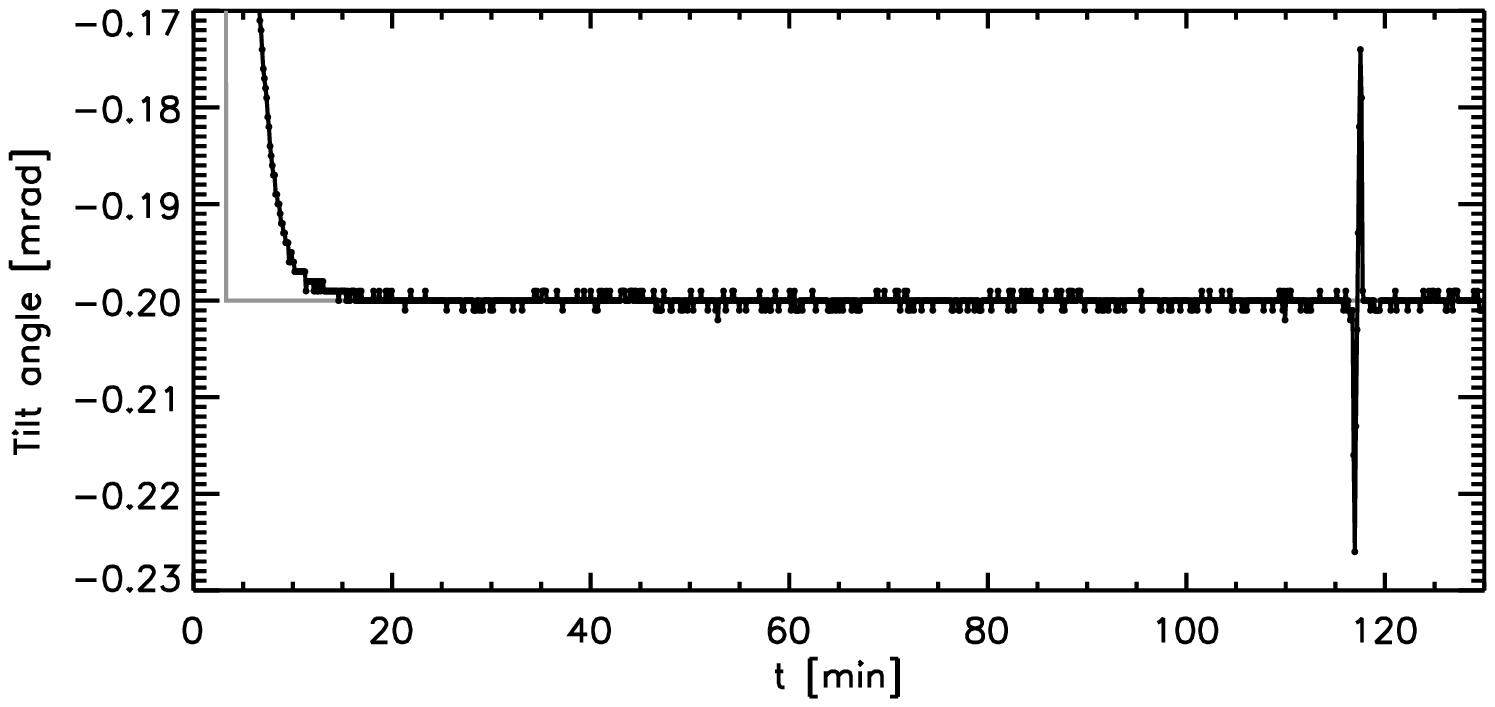}
\end{center}
\caption[sq] {\label{tilt} \em Top: Motor positions of both tripod
actuators that correct tilt perturbations of the whole experimental
setup. Both actuators show a periodic alternation with a period of
24 hours due to thermal expansion of experimental parts reflecting
the temperature dependence of the day/night cycle. The corrections
by the actuators in this example correspond to a tilt compensation
of up to $78\,\mu rad$. Due to this compensation, the residual
deviation from the \textit{set-point} of the tilt is only
$\Delta\alpha=\pm1\,\mu rad$ over many days as seen from the
characteristic noise
level of the bottom graph.\\
Bottom: The signal of one axis of the \textit{NIVEL} inclination
sensor is shown. The whole setup was tilted by $0.2\,mrad$ at
$t=3\,min$ using the tilt control (\textit{proportional}
regulation). The grey line represents the \textit{set-point} value
and the black data the actual measured tilt being adjusted by the
tripod actuators. The tilt of the whole experimental setup was
disturbed for two minutes at $t=116\,min$ by the xy-scan to measure
the particle density profile. There, the camera was displaced by
$\pm 5\,mm$.}\vspace{2cm}
\end{figure}
The experimental setup is exposed to variations of horizontal tilt.
Even small perturbations influence the sample stability. This causes
problems, when experimental arrangements on the experimental plate
need to be changed. Moving parts like e.g. the camera, which
displaces only by $\pm 5\,mm$ horizontally during a density profile
scan, tilt the whole setup significantly ($\pm 25\,\mu rad$, see
bottom Figure \ref{tilt} at $t=116\,min$, scans are switched off
during data acquisition). Another serious disturbance are slow
temperature variations over days. Air conditioning keeps the
surrounding room temperature stable to an accuracy of $\pm1°C$
(temperature sensor of \textit{NIVEL}),
but this is still not sufficient to suppress material expansion.\\
To compensate these disturbances and to achieve a stable horizontal
sample position, the whole experimental setup was installed on a
heavy aluminum base plate standing on a tilt controlled tripod. Two
stands of the tripod are heavy duty
actuators\footnote{\textit{Physik Instrumente}, DC Mike M-235.5DG,
Maximum Load: 120N, Minimum incremental motion: $ 100\,nm$.}, and
the third stand is a rigid pike. A \textit{NIVEL} inclination
sensor\footnote{\textit{Leica}, Nivel20} is mounted on the aluminum
base plate and measures the actual tilt of the setup every six
seconds in x- and y- direction. This values are used as the
\textit{process variables} of the tilt control. \mbox{A
\textit{proportional} regulation} mechanism separately controls the
x- and y-axis of the base plate by adjustment of the tripod
actuators.\\ The top graph of Figure \ref{tilt} shows how strongly
the ambient temperature influences the inclination: the curves show
the \textit{correction variables} (i.e. the positions of the tripod
actuators) compensating tilt variations of more than $78\,\mu rad$
\footnote{This value is calculated from the geometry of the tripod
and the maximum corrections of the actuator positions.}. This
illustrates the necessity of the tilt compensation to acquire data
over several days. An oscillation with a 24 hours period is found in
both \textit{correction variables} of the tilt control reflecting
the non
negligible day-night cycle of the room temperature.\\
The onset of the bottom graph in Figure \ref{tilt} shows an
adjustment of the setup tilt by $200\,\mu rad$ performed by the tilt
control. In approximately 15 minutes the feed back loop regulates
the deviation down to the noise level ($\Delta\alpha=\pm 1\,\mu
rad$) of the \textit{NIVEL} inclination sensor. This tilt adjustment
is a typical step size to correct the density profile at the
beginning of a sample treatment. The step size is decreased when the
sample becomes flatter. An automatization of this correction is not
advisable, because it is not possible to extract a reasonable
\textit{process variable} from the profile scans. Furthermore, after
a change in tilt the system has to equilibrate for at least one day
before another correction can be applied.\\
The \textit{NIVEL} output signals are the most sensitive measures
for inclination of the setup, in particular more sensitive than the
particle density profiles in the cell (see \mbox{section
\ref{density prof})}. Only therefore it is possible to implement a
feedback loop for inclination as the particle profile is susceptible
to changes in
tilt less than $\pm 20\,\mu rad$.\\
\subsection{\label{density prof}Flatness of the interface}
\begin{figure}[b!]
\begin{center}
\includegraphics[width=\columnwidth]{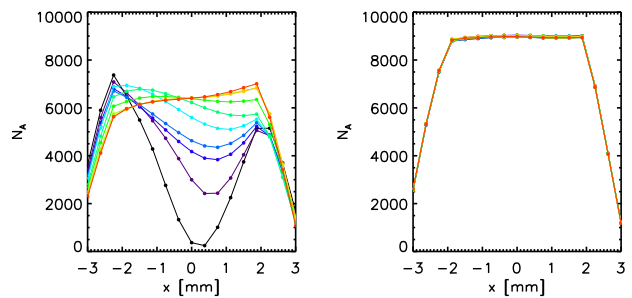}
\end{center}
\caption[sq] {\label{xyscan} \em Left: Scans of particle numbers
$N_A$ per field of view are shown (one component sample), scanned in
x-direction in the first days of sample treatment. Time between two
scans is 20 hours starting with the black curve and ending with the
orange one. The concave droplet is flattened using the number
density regulation over eight days. The inclination was changed
using the tilt control by lifting the positive x-direction by
$500\,\mu rad$. Right: Ten almost identical scans
after several weeks of equilibration and treatment. Time between two scans is two hours.\\
The curves steeply drop at $\pm2\,mm$ where the edge of the cell
enters the field of view. The interaction strength was
$\Gamma=147$.}
\end{figure}
After a sample is placed into the experimental setup, it usually
takes several weeks of treatment before the sample can be considered
as equilibrated and flat. The treatment strategies and stabilization
problems of the sample profiles are discussed now.\\
When the suspension is filled into the glass sample cell using a
conventional $1\,ml$ syringe, the particles sediment to the
water-air interface on the timescale of minutes. Usually at the
beginning of sample treatment, the density profiles across the
sample cell (x- and y-direction) are very inhomogeneous as shown for
the x-direction representatively in the left graph of \mbox{Figure
\ref{xyscan}.} Every two hours a profile scan is performed to track
the development of the sample profiles (left graph displays scans
every 20 hours). The scans show that the sample is less dense in the
center
due to a concave interface and not symmetric.\\
To equilibrate the density and flatten the interface, the density
control (see \mbox{section \ref{densityreg}}) is used to correct the
number of big particles towards lower values over many days or
weeks. Using the tilt control the inclination of the setup is
adjusted to reach a horizontal interface and therefore symmetric
density profiles. Only $200\,\mu rad$ per day and per axis are
corrected since this is the timescale of profile equilibration
\footnote{Here, 'equilibration' does not mean that the density
profile is flat. It only means that the profiles are stable over
time. Density profiles equilibrate faster for low values of
$\Gamma$, but regulation is more stable for high interaction
strengths $\Gamma$.}.\\ The left graph of Figure \ref{xyscan} shows
x-scans of the same sample several weeks later. The profiles are
symmetric and stable over time.\\
\subsection{Equilibration of the binary monolayer\label{prepbehavior}}
Equilibration and treatment of a binary mixture is different to that
of a one-component sample. The binary system reacts much slower to a
change in the interface curvature performed by the density control.
Therefore, the feedback loop parameters of the density control have
to be chosen smaller by a factor of two compared to the
one-component case. The reason for this slower reaction is the
smaller mass per area of the binary monolayer which is only $\approx
35\%$ of the mass per area of the one-component system
\footnote{Typical numbers of $N_A=2000$ big particles and $N_B=1000$
small particles in the field of view for a binary mixture is
compared with $N_A=9000$ big particles in the field of view in the
one-component case.}. Therefore, the binary mixture is less affected
by the 'particle pressure' generated by the downhill slope
gravitational forces at the curved interface. This 'particle
pressure' is indirectly used by the density regulation to control
the density
in the field of view in the center.\\
Furthermore, the glassy flow dynamics of the binary mixture is
different compared to a crystalline one-component system: the
presence of small particles decreases the effective 2D system
viscosity and therefore changes the time for equilibration.\\
\section{Data acquisition}
After typically several weeks of treatment, the sample properties
are sufficiently stable for data acquisition. Then, images are
processed and coordinates are extracted as described in section
\ref{image}. The coordinates are tracked over time \textit{in situ}
to obtain the trajectories of each particle. These coordinates are
stored on the hard disc of the controlling computer in data blocks
each storing $1000$ snapshots. Every data block consists of four
columns: x-coordinate, y-coordinate, recording time, $\pm$ particle
label.\\
The particle label additionally contains the information of the
species. It is negative for small and positive for big particles.
\subsection{'Multiple $\tau$' time steps \label{multiple}}
For investigations of the binary glassy system, data sets have to be
stored over many days when investigating the sample at strong
interaction due to the long relaxation times. Observing typically
$3000$ particles in the field of view, this would exceed storage
volume if the time interval $\tau$ between snapshots was constant,
e.g. $\tau=1\,sec$. Furthermore, the quantities of interest, e.g.
mean square displacements, are usually analyzed on a logarithmic
time scale, and therefore it is not necessary to have the same high
repetition rate of snapshots for the whole time of acquisition.\\
Therefore, the time step $\tau$ between particle snapshots is
doubled every $1000$ snapshots typically starting with
$\tau=0.5\,sec$. However, this requires to track all particles
\textit{in situ} because the time gaps between the stored snapshots
become too long to track particles after data acquisition. In this
way the size of data sets is limited to a few hundred megabytes,
which is still convenient for the subsequent data evaluation on
conventional personal computers.
\subsection{Drift compensation}
\begin{figure}[t]
\begin{center}
\includegraphics[width=0.49\columnwidth]{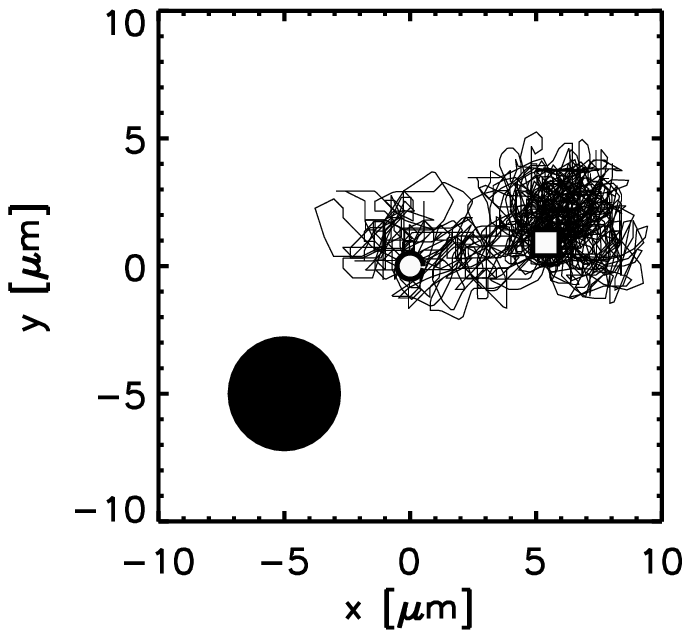}
\includegraphics[width=0.49\columnwidth]{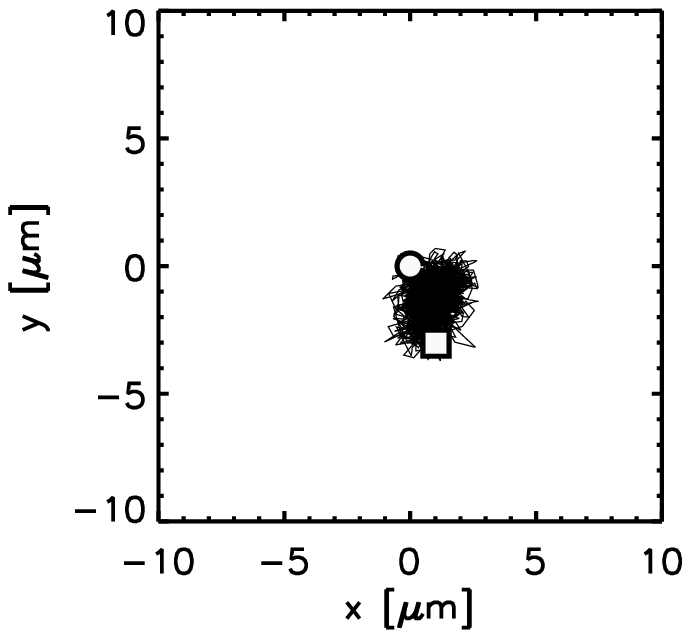}
\end{center}
\caption[sq] {\label{drift compensation} \em Motor drift
compensation for x- and y-direction is plotted left and the residual
overall drift in the sample is plotted right. The starting points
(circle) at the graphs origin and the ending points (square) are
highlighted. The interaction strength was $\Gamma=527$ and the
sample had a relative concentration of $\xi=29\%$. The recording
time was 49 hours. The dark disc for comparison has the diameter of
a big \mbox{particle $(d=4.5 \mu m)$.}}
\end{figure}
In the ideal case a flat and equilibrated sample shows no net drift
of the particles in the field of view. However, at low $\Gamma$ some
small net drift is often inevitable causing loss of particle
information at the edges of the field of view. To compensate a
possible drift, the actuators of the microscope optics mount (see
Figure \ref{setup}) are compensating the average particle
displacement during data acquisition. The drift is calculated and
averaged from the trajectories of all particles in the field of
view, and then x- and y-velocities of the actuators are separately
adjusted by a feedback loop.\\ The total trajectory of the
compensating actuators and therefore of the whole optics mount is
shown in Figure \ref{drift compensation} (left). The displacement
during two days is comparable to the size of a big particle drawn in
the same graph for comparison. This shows how quiescent the
monolayer is for high interaction strengths $\Gamma$. At lower
interaction strengths, the net drift can be larger up to $\approx 10
\mu m/hour$ but is precisely compensated. The accuracy of the
compensation is demonstrated in the right graph of \mbox{Figure
\ref{drift compensation}.} The residual average displacement of all
measured particles is plotted, and it only extends over a few
microns. This residual drift of all measured
particles is subtracted before data evaluation.\\
The drift compensation and additionally the subtraction of the
residual drift after measurement is essential for the investigation
of the system dynamics. Average displacements at high interaction
strengths $\Gamma$ can be much smaller than the drift. Slight drift
deviations strongly matter.\\
Note, that by applying the drift compensation it is assumed that the
average displacement originating from the system is zero in the
field of view, and only perturbations are compensated. However, it
cannot be excluded that inherent information of the system are
obscured, e.g. long wavelength fluctuations as expected for 2D
systems
(\textit{Mermin-Wagner-Hohenberg} theorem \cite{mermin1,mermin2}) or large-scale dynamical heterogeneities.\\
\section{Conclusion}
This paper reports the detailed description of a method to produce a
model system for 2D physics consisting of a monolayer of
super-paramagnetic particles at a water-air interface of a pending
water droplet. The most important advantage of this experimental
geometry is the fact that all particles are uniformly free to
diffuse in two dimensions unhindered by any kind of substrate. From
image processing of video microscopy pictures several thousand
trajectories are obtained providing the complete phase-space
information as well as detailed local features. The technical
equipment is presented and the requirements for sufficient sample
preparation and stability are explained. The inclination of the
setup is controlled to an accuracy of $1\,\mu rad$ which is
necessary for sufficient sample stability. Different interacting
regulation mechanisms control the water volume of the sample cell to
produce a truly two dimensional and flat interface. This was
demonstrated by the constant density profiles across the sample
which are flat and stable over months. The fluctuations of the
particle number density in the field of view are reduced to less
than $\pm 1.3\,\%$. Particle drift is of the order of a few microns
and a small residual collective drift is compensated by active
camera tracking. Furthermore, this method allows manipulation of the
sample by optical tweezers from top as the observation and
regulation is performed by the microscope optic from below the
sample.\\ These technical specifications of the setup provide the
necessary sample quality and stability to investigate the model
system of a 2D crystal or a 2D glass former respectively.
\begin{acknowledgments}
This work was supported by the Deutsche Forschungsgemeinschaft
Sonderforschungsbereich 513 project B6, Sonderforschungsbereich
Transregio 6 project C2 and C4 and the International Research and
Training Group "Soft Condensed Matter of Model Systems" project A7.
Pioneering contributions to the 2D setup from R. Lenke and K. Zahn
are gratefully acknowledged. We thank A. Wille, G. Haller, U.
Gasser, C. Eisenmann, C. Maas, R. Hund, and H. König for their
contributions to the development of the setup and for discussion.
\end{acknowledgments}

\end{document}